\documentclass[5p,number,sort,compress]{elsarticle}
\usepackage[utf8]{inputenc}
\usepackage[T1]{fontenc}
\usepackage{amsmath}

\usepackage{txfonts}

\usepackage{amssymb}
\usepackage{booktabs}
\usepackage[hyphens]{url}
\usepackage{hyperref}
\usepackage{microtype}
\usepackage{lineno}

\begin{document}

\begin{frontmatter}
  \title{Detecting changes in maps of gamma spectra with Kolmogorov--Smirnov
    tests}
  \journal{Nuclear Instruments and Methods}

  \author[cmu]{Alex Reinhart\corref{cor}}
  \ead{areinhar@stat.cmu.edu}
  \cortext[cor]{Corresponding author}

  \author[cmu]{Valérie Ventura}
  \author[arl]{Alex Athey}

  \address[cmu]{Department of Statistics, Carnegie Mellon University, Pittsburgh,
    PA 15213, USA}
  \address[arl]{Applied Research Laboratories, The University of Texas at Austin,
    Austin, TX 78713, USA}

  \date{\today}

  \begin{abstract}
    Various security, regulatory, and consequence management agencies are
    interested in continuously monitoring wide areas for unexpected changes in
    radioactivity. Existing detection systems are designed to search for
    radioactive sources but are not suited to repeat mapping and change
    detection. Using a set of daily spectral observations collected at the
    Pickle Research Campus, we improved on the prior Spectral Comparison Ratio
    Anomaly Mapping (\textsc{scram}) algorithm and developed a new method based
    on two-sample Kolmogorov--Smirnov tests to detect sudden spectral
    changes. We also designed simulations and visualizations of statistical
    power to compare methods and guide deployment scenarios.
  \end{abstract}


\end{frontmatter}

\section{Introduction}

The threat of dirty bombs and lost or stolen radioactive sources has prompted
the development of a variety of systems to detect and identify radioactive
materials, ranging from van-mounted gamma imaging systems to backpack-based
search systems. These systems are typically designed for border checkpoints,
source search, or source identification, but not for the continuous monitoring
of a wide area. Here we investigate detecting changes in radiation spectra over
a wide area, such as a city, stadium, campus, or large public event. Our goal is
to develop an automated mobile sensor system which could monitor radiation
spectra over time and detect sudden changes that might indicate the introduction
of a radioactive source.

The fastest and most sensitive existing method for mapping radiation over a wide
area is a low-altitude helicopter survey; the Department of Homeland Security
has funded several helicopter surveys of large cities, producing maps used as a
baseline for emergency response
plans \citep{Wasiolek:2007vn,RSL:2011}. However, the high cost of operating
helicopters makes it impractical to use them to monitor a city over a long
period of time.

Previous ground-based efforts have focused on source search: traveling through a
city and locating a lost or stolen source when no prior radiological survey is
available. Because the natural background radiation varies from place to place
due to geology and construction materials, these systems must separate natural
variation from variation due to a target radioactive source, usually by assuming
that natural variation is much smaller than that caused by the target source, or
by examining only the energy ranges typical of target
sources \citep{Pfund:2007}. This limits their sensitivity---a small target
source may hide among the variation in the natural background, or may emit at
energies not chosen for targeting.

A long-term radiation surveillance system could avoid this problem by comparing
newly recorded spectra with previous observations \emph{at the same
  location}. For example, we previously developed the \textsc{scram} algorithm,
which is meant to be used with mobile detectors that repeatedly patrol the same
area, recording spectra with timestamps and GPS locations
\citep{Reinhart:2014}. The map is divided into grid cells and each cell's
spectrum is compared to previous observations in the same cell. \textsc{scram}
does not discriminate between source types, instead using its knowledge of the
background spectrum to know what spectra are expected.

However, \textsc{scram} has shortcomings: it downsamples energy spectra into
only eight bins, which potentially limits sensitivity to small or distant
sources, and it requires several repeat mappings of the same area to estimate
accurately the covariance structure between energy bins.

We propose a new method based on Kolmogorov--Smirnov tests which, like
\textsc{scram}, can detect any spectral changes regardless of type, but requires
no covariance estimates and no downsampling, and hence can work with less
background data. This method is simpler, has higher power and provides better
source localization than \textsc{scram}. To guide detector deployment, we
present simulations and visualizations of statistical power which allow
operators to find areas of vulnerability.

\section{Data}
\label{data}

\begin{figure}
  \centering
  \includegraphics[width=\columnwidth]{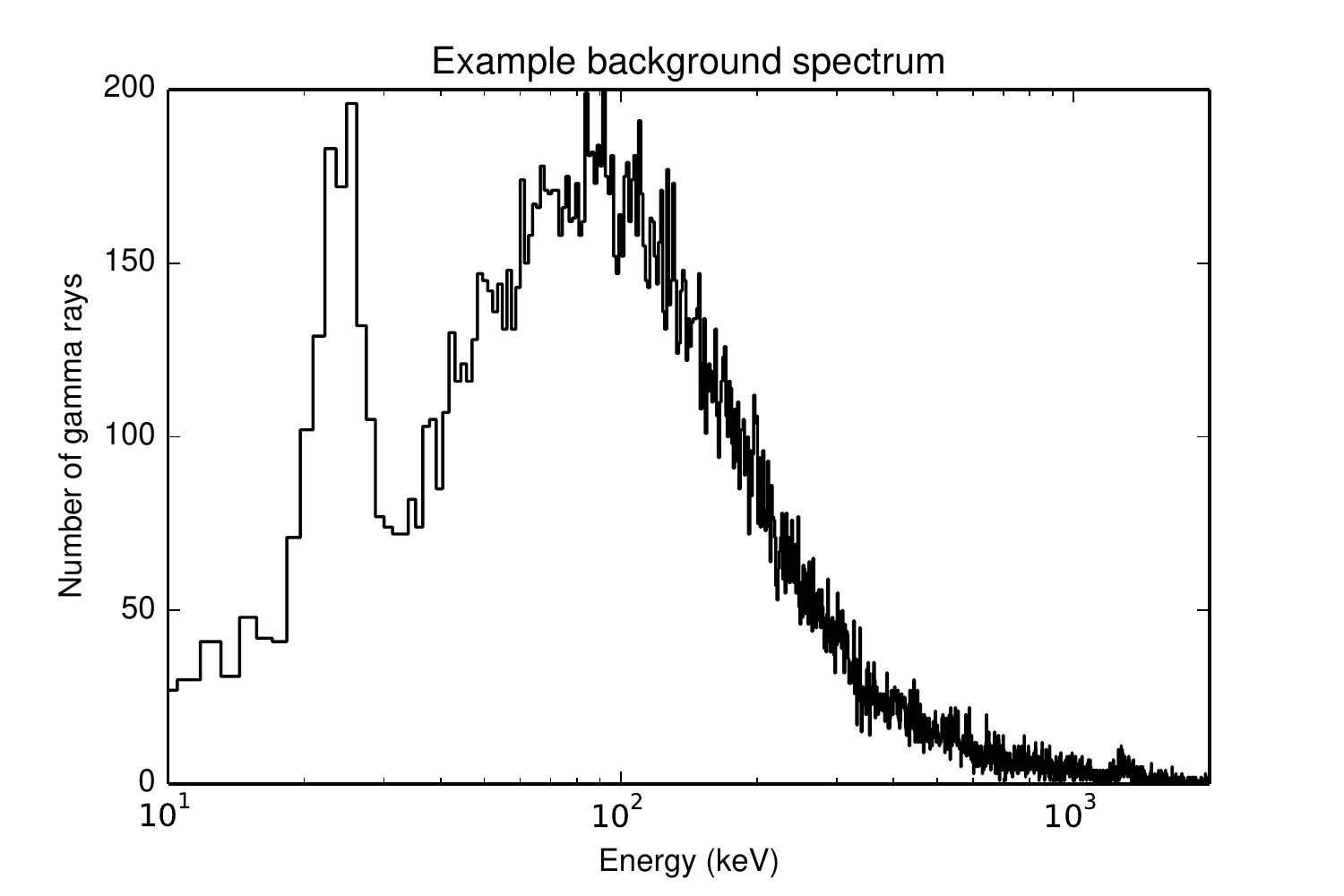}
  \caption{A typical background radiation spectrum at Pickle Research Campus,
    comprising 32,173 gamma rays observed over several hours. Energy in
    kiloelectronvolts, in 4,096 bins, is shown on a logarithmic scale.}
  \label{arl-bg}
\end{figure}

We collected our data using a \(2 \times 2\) inch Bridgeport Instruments cesium
iodide spectrometer, a laptop, and a GPS unit. The spectrometer continuously
recorded gamma rays and produced a 4,096-bin spectral histogram every two
seconds; the laptop then recorded the histogram, time, and location. In typical
conditions, 80--120 gamma rays were observed per histogram. An example spectrum,
consisting of typical background gamma rays and summed over several hours, is
shown in Fig.~\ref{arl-bg}, while Fig.~\ref{cs137} shows a sample taken near a
radioactive cesium-137 source.

\begin{figure}
  \centering
  \includegraphics[width=0.8\columnwidth]{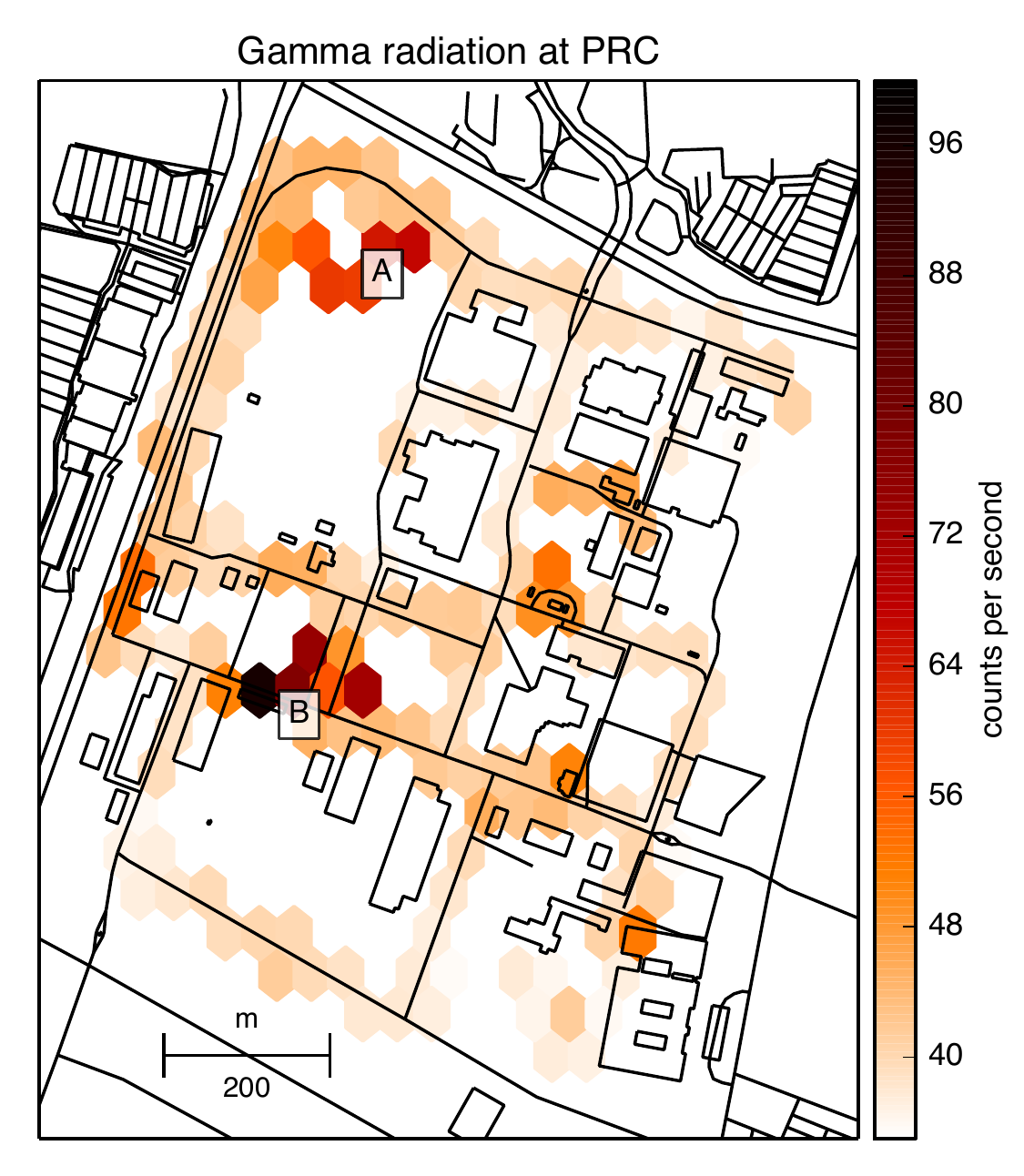}
  \caption{A map of Pickle Research Campus with total gamma counts per second
    overlaid; counts are averaged over one month of data collection. Areas of
    elevated background include the radioactive materials storage facility at
    the northwest corner (A) and a cluster of large brick buildings near central
    campus (B). Figure reprinted from \citep{Reinhart:2014}.}
  \label{prc-cps}
\end{figure}

The dataset consists of once- or twice-daily drives through Pickle Research
Campus (PRC) in the months of July and August 2012. The spectrometer and GPS
unit were loaded onto a golf cart and driven around campus for roughly half an
hour. Various spectral features at PRC, such as slightly-radioactive brick
buildings and a radiological waste storage site, cause the area to have total
background radiation levels which vary in space by about a factor of three; this
variation is shown in Fig.~\ref{prc-cps}. Cumulatively, the data includes
roughly 18 hours of observations taken over 41 drives through campus on 30
different days.

In the course of our analysis we discovered that the dataset is contaminated: we
used our Kolmogorov--Smirnov anomaly detection algorithm (Section \ref{ksad}) to
compare each day to the previous day and identified days with unusual spectral
differences, the largest of which is likely due to a downpour of 7~cm of rain
the previous evening; rain can cause large variation in background spectra
\citep{Shebell:1996hw,Livesay:2014eo}. In the rest of our analysis we excluded
this day. (This was the largest rain event during the dry Texas summer, and the
only to cause a noticeable anomaly.) This ensures our estimates of false
positive rates (Section \ref{ksad}) do not contain true positives; we'll instead
use simulated sources of known size and location to test our algorithms. Future
work may be able to account for rain-induced spectral changes using a model to
relate precipitation rate and radon progeny deposited by rain
\citep{Mercier:2009gv}.

\section{Approach}

To detect radioactive sources, two things are required. The first is a way to
account for the natural spatial variation in background spectra (Section
\ref{bgmodel}). The second is an anomaly detection algorithm which compares the
background model with new observations and tests for statistically significant
differences, producing a map of anomalous regions (Sections \ref{scramfix} and
\ref{ksad}). Global false discovery control is essential to make the system
practical, and the power of the procedure needs to be established for target
radioactive sources.

The anomaly detection algorithm should use only the shape of the spectrum, not
the total count rate, since observed count rates will vary widely depending on
the detector, and a wide area monitoring system may use different sizes of
detectors mounted on different vehicles at different heights. Also, like
\textsc{scram}, our anomaly detection algorithm does not attempt to discriminate
between benign and threatening anomalies, instead searching for any spectral
change; users who need to search for specific sources can check anomalies using
a source identification algorithm to locate specific isotopes
\citep{Boardman:2013fc,Boardman:2013cs}.

\subsection{Background Radiation Model}
\label{bgmodel}

The background spectrum arises from natural deposits of uranium, thorium, radon,
and their decay products; these elements are more common in certain kinds of
rock, such as granite, and hence the spectrum varies with local geology. This
variation is usually smooth, though sudden changes in geology (such as a
man-made granite wall) can cause sharp background changes. Similarly, different
concrete mixes can have different mixtures of naturally radioactive materials,
causing variation in the background emitted by concrete structures
\citep{Ryan:2011wh}. To avoid comparing new data to a heterogenous background,
we divide the map into spatial cells and aggregate each cell's data over several
days to provide a background estimate.


\begin{figure}
  \centering
  \includegraphics[width=\columnwidth]{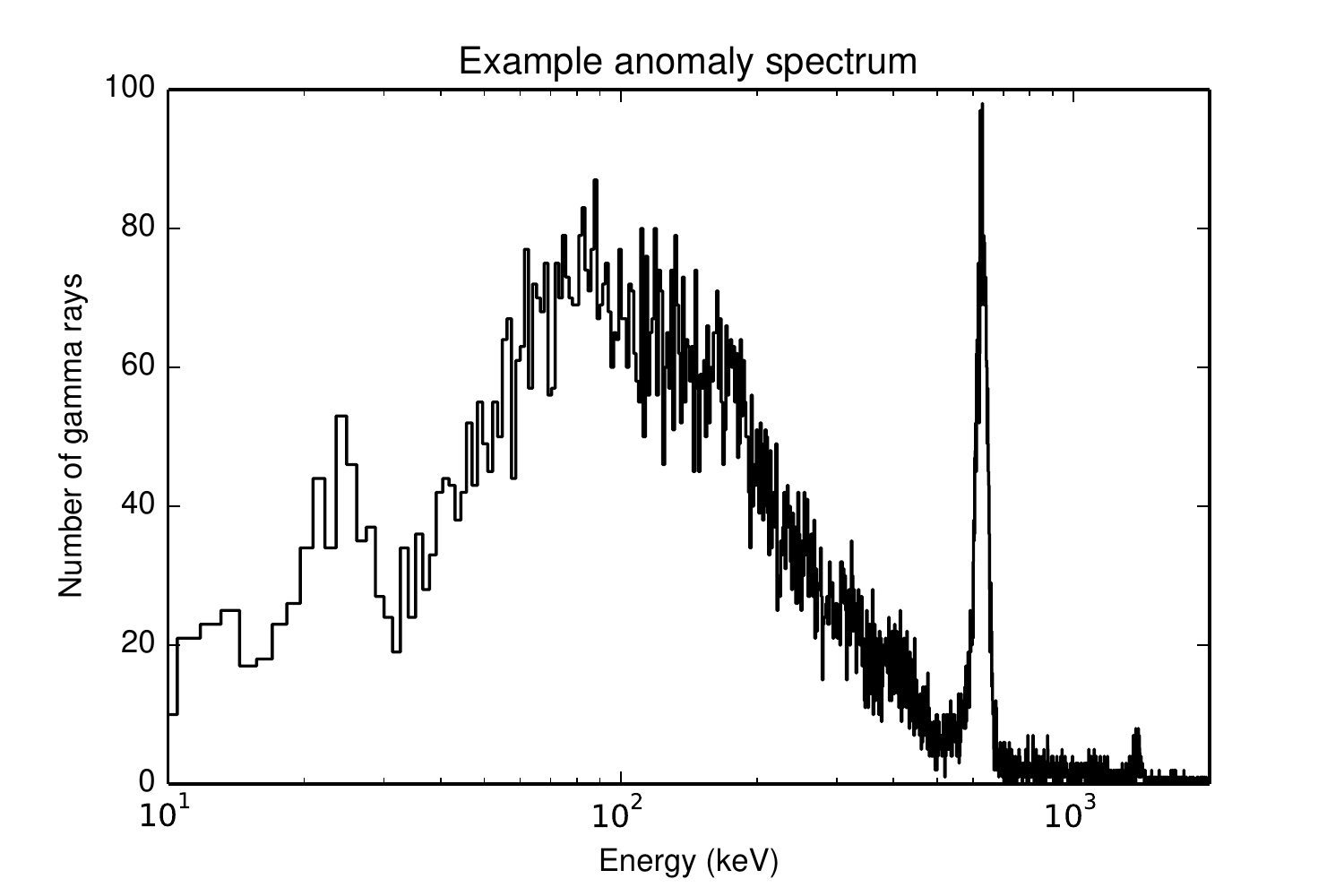}
  \caption{Spectrum recorded near a sample of radioactive cesium-137. The sharp
    peak on the right side of the plot, which stands out from the normal
    background, is the characteristic 662 keV gamma ray emitted in the decay of
    cesium-137.}
  \label{cs137}
\end{figure}

\subsection{Spectral Comparison Ratio Anomaly Mapping (\textsc{scram})}
\label{scramfix}

The \textsc{scram} procedure was our initial attempt to perform anomaly
detection \citep{Reinhart:2014}. It is based on the method of spectral
comparison ratios of \cite{Pfund:2007} and proceeds as follows: new spectra to
be tested are aggregated in \(250 \times 250\) meter spatial cells (there are
typically 12--15 such cells per day for the data analyzed here), downsampled
from 4,096 energy bins to 8, and a $\chi^2$-like anomaly statistic $D^2$
calculated in each cell by comparing the 8-bin spectrum to spectra recorded on
previous days in the same cell.  Large values of $D^2$ relative to a reference
distribution indicate elevated radioactive emissions in the corresponding cells.

\textsc{scram} makes various simplifying assumptions: count rates have a
constant variance-to-mean ratio and background spectra are treated as
exact. These assumptions imply that the \(\chi^2\) distribution assumption
underlying the test statistic $D^2$ does not hold; in particular, the
distribution has a heavier tail than a \(\chi^2\) \citep{Reinhart:2014}. We
instead estimated the reference distribution of $D^2$ with a histogram of values
of $D^2$ calculated for each of the 30 days of recorded data.  This resulted in
399 observed test statistics, 12--15 per day, assumed to be from the null
hypothesis that no genuine spectral difference exists. The 99th percentile of
these values was \(D^2 = 83\), which was then used as the cutoff for an \(\alpha
= 0.01\) level test.

We consider a slightly modified version in this paper. Indeed, this cutoff value
is not only variable, since it is estimated from a small sample, but it is also
biased, since the dataset contains true positives, as mentioned
earlier---rain-induced spectral changes produced anomalies across the entire map
for one day. These outlying points inflated the \(D^2\) cutoff.  We
excluded this day of observations from the dataset rather than attempting to
isolate specific affected cells. We then repeated the procedure and calculated a
parametric rather than a nonparametric cutoff: we fit a gamma distribution with
probability density function
\[
f(x) = \frac{\beta^\alpha}{\Gamma(\alpha)} x^{\alpha-1} e^{-\beta x}.
\]
to the resulting 387 test statistics by maximum likelihood, and used its 99th
percentile, \(D^2 = 39.7\), as the cutoff.  The gamma distribution includes the
\(\chi^2\) distribution as a special case but is more flexible, allowing in
particular for the heavier tails caused by the assumption violations
\citep{Reinhart:2014}.  This parametric cutoff may be biased if the gamma
reference distribution is not perfectly adequate, but it is also less sensitive
to outliers and has lower estimation variance than the empirical cutoff of
\citep{Reinhart:2014}.  This lower rejection threshold has substantial impact on
our power comparisons (see Section~\ref{power-results}); the revised
\textsc{scram} method is more sensitive than originally proposed.

Note that the nonparametric cutoff of \citep{Reinhart:2014} and the parametric
cutoff calculated here are estimated from data, so they are variable. There is
no standard analytical formula for the confidence interval of this estimate, so
we used bootstrapping to estimate the uncertainty in this fit, using the
following procedure \citep{DiCiccio:1996cl}:
\begin{enumerate}
  \item Re-sample at random and with replacement from the 387 test statistics.
  \item Fit the gamma distribution to this new sample and calculate its 99th
    percentile.
  \item Repeat 10,000 times.
\end{enumerate}
The quantiles of the 10,000 repeat values of the cutoff are used to set the
95\% basic bootstrap confidence interval of \([33.5, 45.2]\), suggesting more
data would be needed to more accurately choose a rejection threshold.

\subsection{Kolmogorov--Smirnov Anomaly Detection}
\label{ksad}

We developed an alternative approach that does not require empirical estimation
of its rejection region, does not reduce its detection power by downsampling
spectra, and does not require repeat days of observation to estimate the
correlation between energy bins. Our approach uses a standard two-sample
Kolmogorov--Smirnov (KS) test.

We again divide space into square cells and compare each cell to prior
data. This approach treats gamma rays as draws from an underlying energy
distribution, with the KS test checking if two samples come from the same
distribution.  Specifically, if \(c_{j1}, c_{j2},\dots, c_{j{n_j}}\) is a list
of energies of \(n_j\) gamma rays from distribution \(j\), we let \(\hat
F_j(x)\) be their empirical cumulative distribution function---the proportion
with energy less than or equal to \(x\):
\[
\hat F_j(x) = \frac{1}{n_j} \sum_{i=1}^{n_j} \mathbf{1}(c_{ji} \leq x),
\]
where \(\mathbf{1} (\cdot)\) is the indicator function, which is \(1\) when its
argument is true and \(0\) otherwise. In our case, \(\hat F_1(x)\) is the
observed distribution of background gamma ray energies, and \(\hat F_2(x)\) the
distribution of a new observation. Then the test statistic is
\[
D = \sup_x \left| \hat F_1(x) - \hat F_2(x) \right|.
\]
When both samples are drawn from the same continuous distribution, \(D\) has a
null distribution which is easily approximated by standard statistical software
packages. When they are not from the same distribution, $D$ tends to be large
and the test rejects the null hypothesis, though as with all tests, its
sensitivity decreases when the sample sizes \(n_1\) and \(n_2\) decrease.

The standard approximations for the KS test statistic null distribution assume
that the data is drawn from a continuous probability distribution, so that no
two observations have exactly the same value. However, our detector creates ties
by discretizing energies into bins, biasing the test's $p$ values to be larger
than expected. We checked this by repeatedly simulating (resampling with
replacement) from a single reference spectrum, then testing the simulated
spectra against the reference spectrum; despite the spectra being draws from the
same distribution, their $p$ values were not uniformly distributed but were
skewed towards one. Biased $p$-values are problematic because they imply that
the false positive rate of the test does not match the chosen significance
level.

This problem could be solved by using a nonparametric permutation test, which
simulates the null distribution of \(D\) by repeatedly permuting the labels on
the data. The gamma energies are gathered into a single list \(c_{11}, c_{12},
\dots, c_{1{n_1}}, c_{21}, c_{22}, \dots, c_{2{n_2}}\) and the following
procedure is run:
\begin{enumerate}
  \item Randomly shuffle the data list. Treat the first \(n_1\) entries as
    samples from distribution 1 and the remaining \(n_2\) entries as samples from
    distribution 2.
  \item Calculate \(D\) with the new samples.
  \item Repeat at least 1,000 times. Calculate the $p$ value as the fraction of
    $D$ values larger than the unshuffled $D$.
\end{enumerate}
The shuffling ensures there is no systematic difference between the two samples,
so we can simulate the distribution of $D$ under the null hypothesis without any
parametric assumptions. This eliminates the bias but requires time-consuming
simulations. Alternatively, \citep{Schoer:1995bm} describes a procedure for
calculating an exact $p$ value in the case of ties, but for large sample sizes,
the calculation is even slower than the permutation test.  We next tested
whether we could avoid these costly adjustments by calibrating the $p$ value
threshold to get an approximate $\alpha = 0.05$ test, and whether this test
would have power equivalent to the permutation test.

We again used a simulation procedure, repeatedly resampling from a background
distribution and injecting gamma rays from a cesium-137 source. First, we ran
the simulation 10,000 times without the injected source, producing a null
distribution, and calculated that 5\% of the KS test's $p$ values were below $p
= 0.067$, so we adopted this as our rejection threshold. Then we ran 1,000
simulations for each of several injected source sizes and compared the power of
this adjusted KS test to the unadjusted permutation test. They were virtually
identical, implying that the KS test with adjusted rejection threshold will
suffice.

Finally, we checked how the bias manifested in real data. For every day of data,
we compared the observations in each spatial cell to the previous day's
background measurements, then calculated the fraction of $p$ values below the
unadjusted $\alpha = 0.05$ threshold. We repeated this procedure for several
cell sizes between \(20 \times 20\) and \(100 \times 100\) meters. Results are
shown in Table~\ref{nulldist}. We found that for \(20 \times 20\)m spatial
cells, the proportion of test statistics below \(\alpha = 0.05\) was \(0.052\),
close to $0.05$. For large \(100 \times 100\)m cells the $p$ values are biased
downwards, with nearly 12\% of $p$ values smaller than the $\alpha = 0.05$
threshold.

This is the opposite of the bias we would expect from ties, and must come from
some other source. To confirm this, we repeated the same simulation procedure
but produced $p$ values using the permutation test. This test does not suffer
from bias caused by ties, but Table~\ref{nulldist} shows that its $p$ values
become even more biased towards zero than the KS test's as the cell size
increases.  This bias occurs because large spatial cells are spectrally
heterogenous, so there are genuine differences to be detected. The permutation
and unadjusted tests both pick this up and report more \(p < 0.05\) results, but
for the latter this shift is partly balanced out by the opposite skew caused by
ties. This suggests that the unadjusted KS test would be more useful in
practice.

For the remainder of this paper we will therefore use the KS test, as it is much
faster to compute, gives results which have good detection power, and has
approximately the correct false positive rate when used with \(20\times 20\)m
spatial cells. Larger cells would have larger sample sizes and greater power,
but would have more false positives and be less able to localize small sources,
so there is a tradeoff. When applied to other datasets with different detector
sensitivity and spatial variability, the optimal cell size may differ.

\begin{table}
  \centering
  \begin{tabular}{r r r r}\toprule
    {Cell size (m)} & {Sample $n$} & {KS \(\alpha\)} & {Permutation \(\alpha\)}\\\midrule
    20 & 4604 & 0.052 & 0.064\\
    40 & 2412 & 0.070 & 0.082 \\
    100 & 949 & 0.116 & 0.133\\\bottomrule
  \end{tabular}
  \caption{For each size of spatial cell, the number of KS tests performed and
    the proportion of test statistics smaller than the nominal \(\alpha = 0.05\)
    level. The permutation test appears to have a higher false positive rate
    than the standard test using the asymptotic null distribution.}
  \label{nulldist}
\end{table}

\subsection{False Discovery Rate Detection}

To produce an anomaly map, the data are divided into a spatial grid, with each
cell's background observation compared to the new data collected within the same
cell, producing a \(p\) value for each cell. To prevent many false detections
due to multiple comparisons, the Benjamini--Hochberg procedure
\citep{Benjamini:1995ws} is used to select a \(p\) value cutoff to achieve a
chosen false discovery rate, defined as the expected proportion of detections
that are false out of all the detections. In this procedure, the cell $p$ values
are sorted into ascending order, \(p_1 \leq p_2 \leq \dots \leq p_m\), and the
significance threshold is set to be the largest $p_i$ such that \(p_i \leq i q /
m\), where \(q\) is the desired false discovery rate. This guarantees that, on
average, the fraction of significant results which are statistical false
positives will be $q$. We set \(q = 0.2\) for our simulations, but $q$ can be
chosen to meet operational needs.

\section{Results}
\label{results}

We built on previous simulation tools \citep{Reinhart:2013wa} to create a set of
power simulations to evaluate the performance of our anomaly detection
algorithm. These simulations inject radioactive sources into observed data and
account for the distance between source and detector, both with the normal
\(1/r^2\) falloff and with exponential attenuation due to gamma absorption in
air, but not for physical obstacles, down-scatter, or shielding between the
source and the detector. In this sense they are idealized; heavily shielded
sources are harder to detect than the simulations suggest, and down-scatter will
alter their spectra.

The count rate $\lambda$ for injected sources was based on the assumption that
\[
\lambda \propto \frac{s}{r^2} \exp(-\mu d),
\]
where \(r\) is the distance to the source in meters, \(s\) is its size in
milliCuries and \(\mu = 0.0100029\, \text{m}^{-1}\) is the attenuation
coefficient for cesium-137's 660 keV gamma rays. The proportionality constant
was determined from an experiment which placed a 0.000844~mCi cesium-137 source
0.05 meters away from our detector, recording 630 counts per second, so that
ultimately, our source simulation model was:
\[
\lambda(d,s) = \frac{s}{0.000844} \cdot 630 \cdot \left( \frac{0.05}{r}
\right)^2 \cdot \exp \left( -\mu (r + 0.05) \right).
\]

To test our anomaly detection algorithm we must specify the anomalies of
interest.  The example in Fig.~\ref{cs137} shows the spectrum recorded near a
sample of radioactive cesium-137; a sharp high-energy peak is apparent. Other
anomalies may be different; iridium-192, for example, appears as several smaller
peaks between 200 and 600 keV mixed in among the large background radiation
smear. Statistical power depends on the type of radioactive source present, but
because the KS test uses the cumulative distribution of gamma rays, it is
sensitive to sharp peaks as well as spread-out differences in the spectrum. We
will use cesium-137 for all our power simulations for consistency.

We also need to specify the background observation period based on the expected
rate of change of the background spectrum. If the background is known to vary
naturally over a period of a few days or weeks, we may use only the most recent
few days to form the background to avoid using an outdated background estimate.

Despite this precaution, the selected background and observation periods may
have genuine spectral differences, for example if rain changes the background
spectrum. If these are statistically significant, then every simulation will
result in a statistically significant result. To prevent this, we do not use
data from the new observation period; instead, we sample with replacement the
matching number of gamma rays from the background period in every spatial cell
and inject the radioactive source into that simulated data. In practice, maps
with anomalies reported everywhere would indicate a global phenomena, such as
environmental change or calibration issue, and would be investigated
accordingly.

\subsection{Minimum Detectable Sources vs.\ Distance}
\label{power-results}

\begin{figure}
  \centering
  \includegraphics[width=\columnwidth]{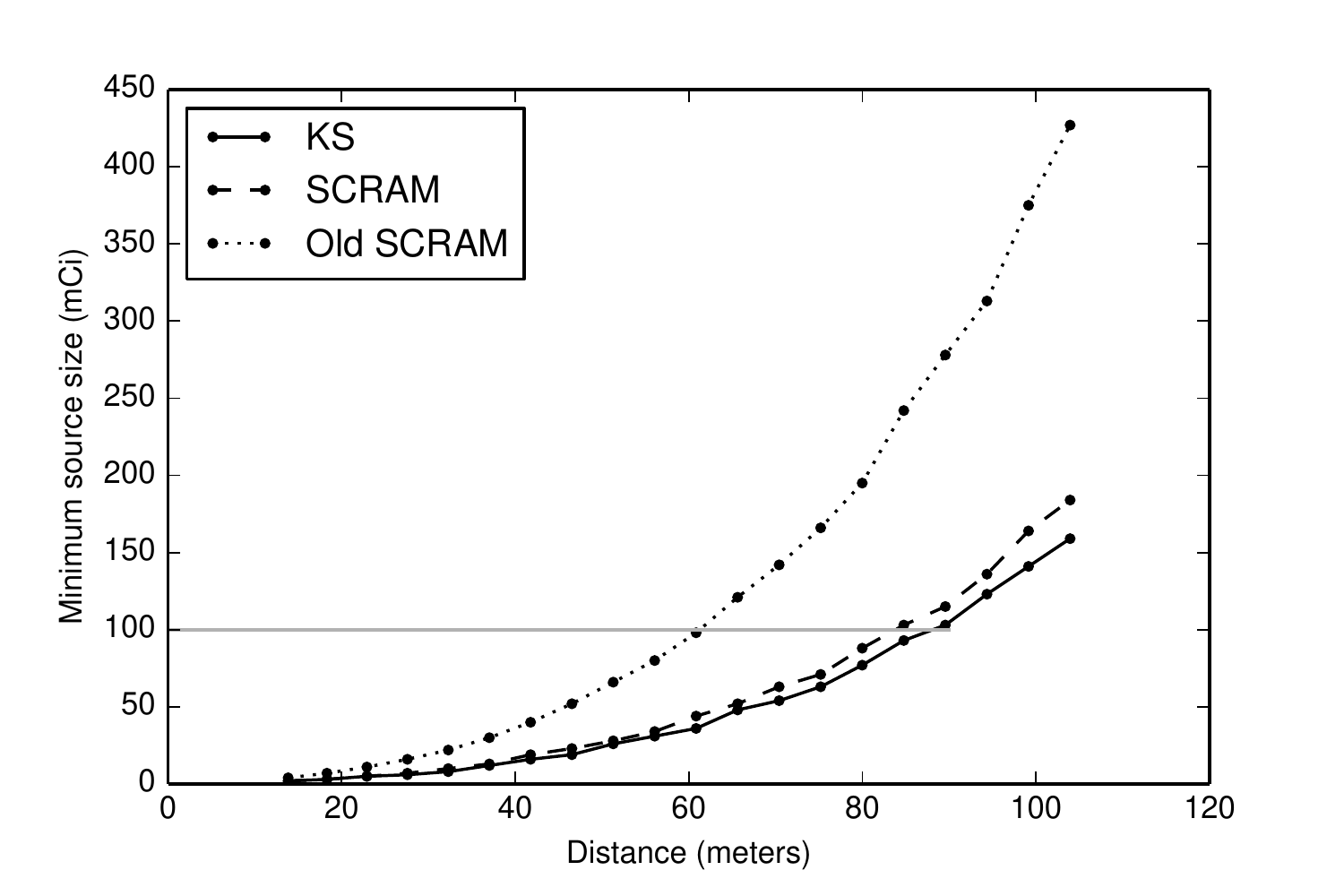}
  \caption{Minimum source sizes detectable by our $2 \times 2$ inch detector
    with 80\% power as a function of distance, compared between KS tests and
    \textsc{scram}. The horizontal line indicates a 100 milliCurie source. KS
    tests are a clear improvement over \textsc{scram}, and the revised
    \textsc{scram} detection thresholds are a major improvement over the old
    version.}
  \label{minimum-detectable}
\end{figure}

To compare the sensitivity of \textsc{scram} and KS tests for detecting small
spectral anomalies, the first simulation determines the minimum radioactive
source size necessary to trigger an alarm (\(p < 0.01\)) with 80\% power, at
varying source distances from the detector. A 20-meter section of road at the
PRC was selected as the test site and simulated cesium-137 sources injected at
various distances from the road. KS tests and \textsc{scram} were run on the
entire aggregated segment of road, using the previous seven days of observations
as background data. (The detection threshold for \textsc{scram} was set by
applying the procedure from Section~\ref{scramfix} to 20-meter cells.)
Fig.~\ref{minimum-detectable} shows the results; it provides the minimum
detectable source at a given distance, and conversely the maximum distance at
which a given source can be detected. For example, the horizontal line indicates
the maximum distances at which a 100 milliCurie source can be detected by each
algorithm at least 80\% of the time. The old \textsc{scram} procedure detects it
at about 65m, while the KS test detects it as far away as 90m, a substantial
improvement in detection power. \textsc{scram} with the revised detection
threshold is also an improvement, performing nearly as well as the KS tests.

\begin{figure}
  \centering
  \includegraphics[width=0.9\columnwidth]{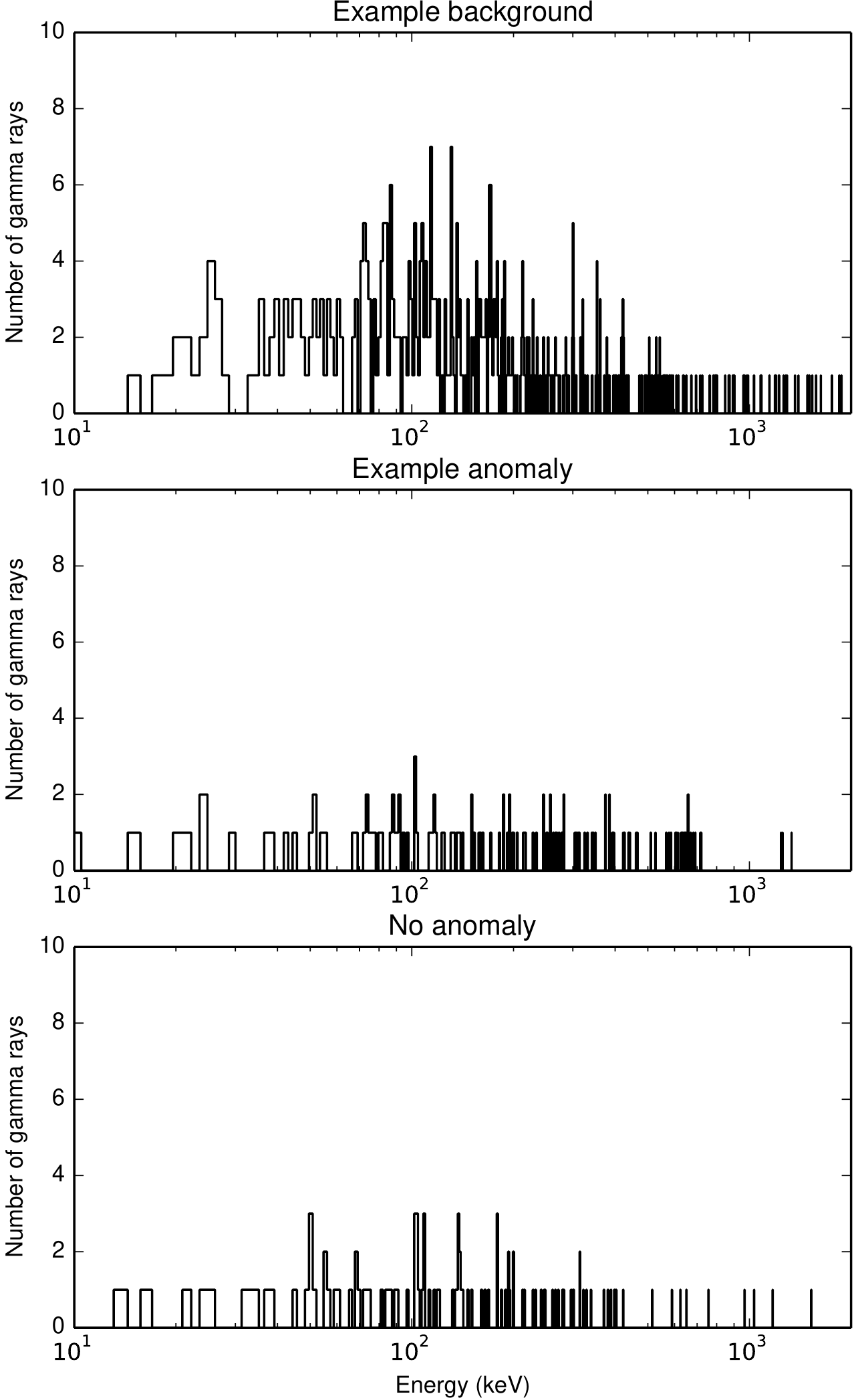}
  \caption{At top, a simulated background spectrum comprising only 504 gamma
    rays drawn from the sample shown in Fig.~\ref{arl-bg}. Center, a simulated
    spectrum comprising 97 background gamma rays and 50 from the cesium-137
    sample shown in Fig.~\ref{cs137}. Bottom, a simulated background of 116
    gamma rays. Despite the small sample, the KS test correctly identifies the
    anomalous spectrum.}
  \label{spectra-compare}
\end{figure}

To illustrate the small amount of data that can be used to detect a source,
Fig.~\ref{spectra-compare} shows a background spectrum (about five seconds of
observations), a simulated contaminated spectrum containing cesium-137 (about
one second of observations), and a simulated spectrum with no
contamination. Despite the small sample sizes, the KS test correctly identifies
a statistically significant difference in the anomalous spectrum. (Recall that
it detects differences in shape, not total number of counts.)

Because the KS test is simpler and more powerful than \textsc{scram}, the
remainder of our power simulations will focus on the KS test. We do not show the
equivalent power maps for \textsc{scram}.

\subsection{Minimum Detectable Source Maps}

We also designed a minimum detectable source simulation tool. The user chooses a
radioactive source, a minimum desired statistical power, a background
observation period, and a new observation day. The simulation grids the data in
the new observation day and sequentially injects a source at each grid location,
recording the minimum source size required to have the desired power of
detection in at least one grid cell. (Again, power is calculated by testing if
the cell would be considered significant after false discovery rate control.)
This produces a map of source sizes: the smallest detectable source size at each
location.

\begin{figure}
  \centering
  \includegraphics[width=0.8\columnwidth]{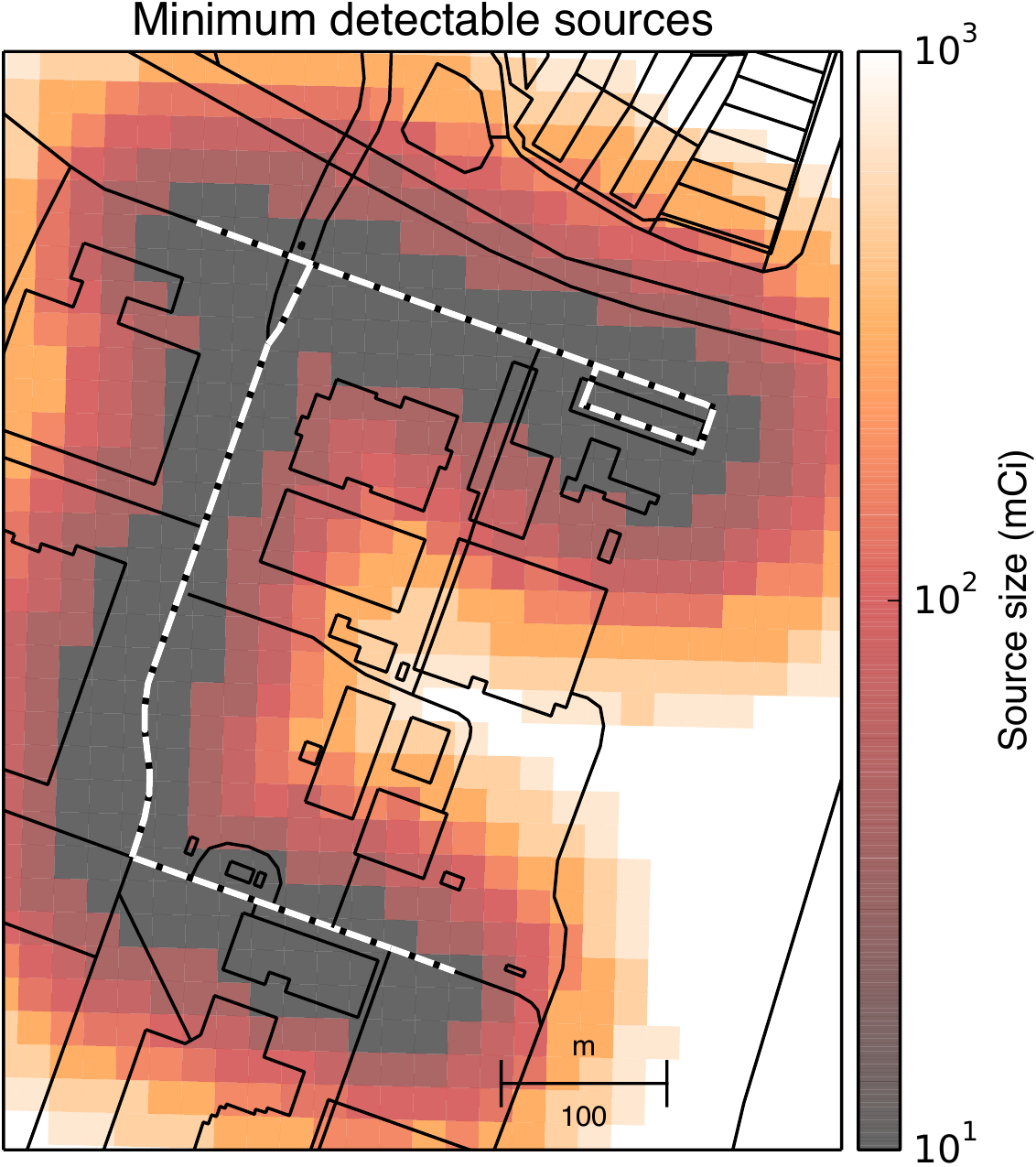}
  \caption{At each location, the smallest cesium-137 source that can be detected
    with 80\% power by a detector which drove through once, using one full week
    of background data. The drive amounted to just five minutes of observations
    along the white dashed route.}
  \label{minimum-map}
\end{figure}

An example map is shown in Fig.~\ref{minimum-map}, which shows that with a week
of daily background observations and only five minutes of new data, we have high
power to detect fairly small radioactive sources along the detector's
route. More distant sources must be much larger to be reliably detected. An
operator could use this map to understand the performance of the system and
evaluate its ability to detect potential threats. This tool could also be
adapted to plan detector routes through an area, identifying routes which
efficiently cover the area of interest.

\subsection{Power Maps for Detecting Chosen Sources}

\begin{figure}
  \centering
  \includegraphics[width=0.8\columnwidth]{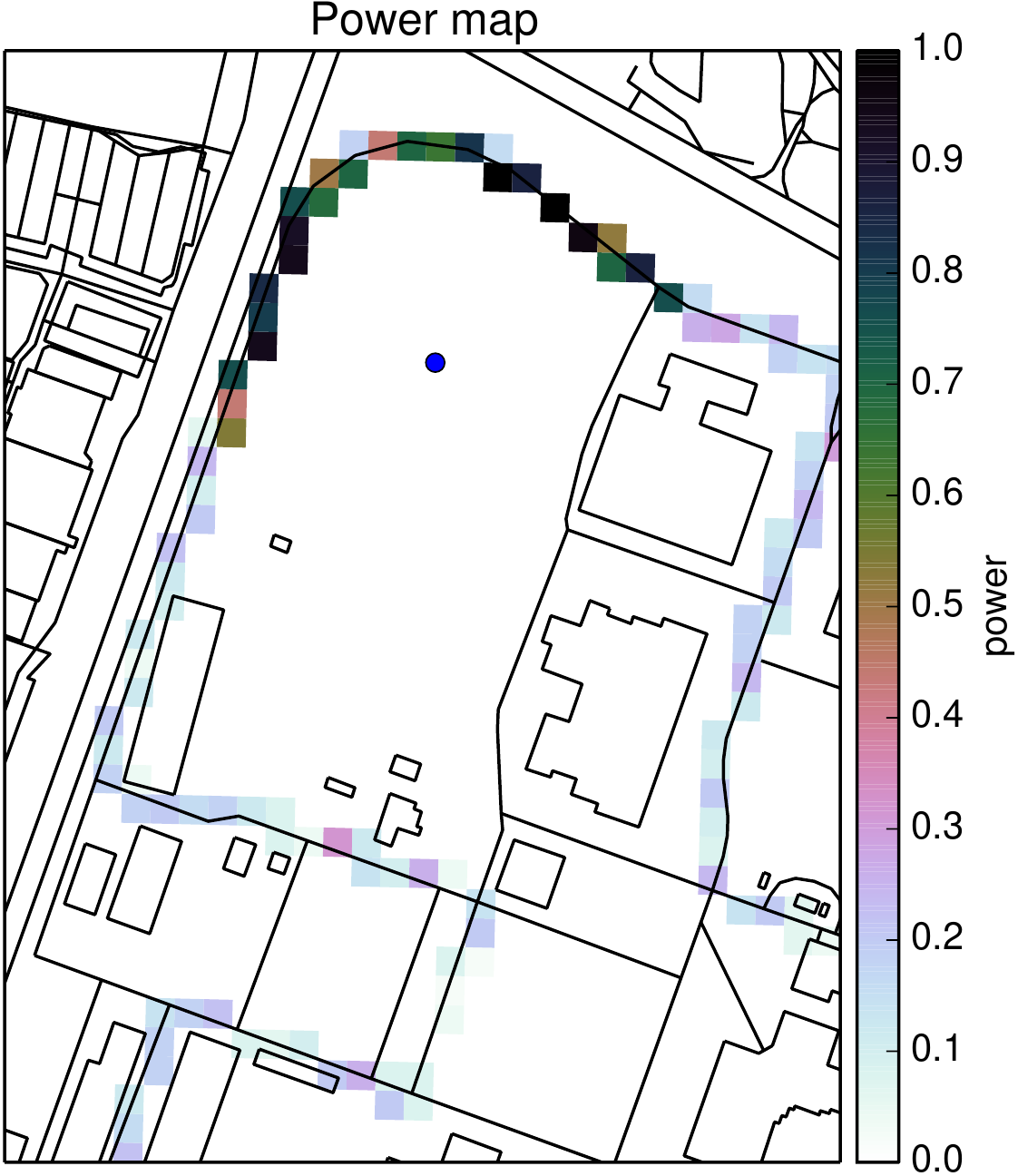}
  \caption{Power to detect a 500 mCi cesium-137 source, injected at the blue
    dot, using one full week of background data (130 minutes of observations)
    and a single pass (22 minutes) of observations of the campus. The map shows
    only the northwest corner of PRC; each grid cell is \(20 \times 20\) meters,
    for a total of 245 cells, so the average cell contains just five seconds of
    observations of the source.}
  \label{power-sim}
\end{figure}

The next simulation we designed was interactive. The user selects a background
observation period and a set of new data. The user then chooses a radioactive
source and size and can click on a map to inject the source at any desired
location in the new data. An anomaly is considered statistically significant if
it passes the Benjamini--Hochberg procedure with a false discovery rate of 0.2,
and the power measures how likely this is to occur in repeated simulations. One
such simulation, using a 500 mCi cesium-137 source, is shown in
Fig.~\ref{power-sim}. The source, placed in an empty field, is easily detected
on the nearby road.

Because the map is interactive---parallelization means the simulations can run
in just a few seconds---the user can explore likely locations for radioactive
sources and determine if the available data is adequate to detect sources at
these locations. If not, detectors could be rerouted to fill the gaps.

A non-interactive version of this map is easy to produce. After choosing a
radioactive source and size, the source is injected into each grid cell, which
is assigned a color proportional to the probability of detecting the source. (A
source is considered to be detected if it results in a statistically significant
anomaly anywhere else on the map, after the Benjamini--Hochberg procedure is
applied.) If there is a specific target source of concern, this map shows where
it would be best detected. We did not include an example here because the map
looks like the inverse of Fig.~\ref{minimum-map}---cells where the minimum
detectable source is small have a higher power to detect the user-specified
source.

\section{Conclusions}

KS tests promise to be a powerful, flexible, and simple alternative to
\textsc{scram} for wide-area mapping and anomaly detection.  More importantly,
we have developed simulation tools that can guide the deployment of our anomaly
detection algorithm in practice, showing that it can be made useful for
real-world applications.

Comparisons against existing mobile source search systems, such as spectral
comparison ratios \citep{Pfund:2007} and the NaI--SS Radiation Search System
\citep{ORTEC}, are difficult. These systems are not designed to monitor a wide
area continuously for a long period of time, and their ability to detect small
and distance sources has been evaluated using different sources and detectors
than we possess. Nonetheless, we believe that our KS procedure is better suited
to long-term radiation surveillance, particularly with the simulation tools we
have developed to aid deployment.

Future work may expand our simulations and tests based on standard procedures
such as those defined in ANSI N42.43--2006 or the Domestic Nuclear Detection
Office's Technical Capability Standard for Vehicle Mounted Mobile Systems. This
would include simulating other common industrial and medical radioisotopes, as
well as simulating shielded sources.

There are also opportunities for future improvement of our anomaly detection
system. We could incorporate methods to detect specific isotopes or ignore known
benign source types, such as common medical radioisotopes, building on previous
work in isotope detection and identification
\citep{Boardman:2013fc,Boardman:2013cs}. Additionally, we have not built a
spatial model of spectra which allows borrowing of information between spatial
cells. Two options are to smooth spectra in space \citep{Tansey:2015} or to use
a spatial false discovery procedure \citep{Benjamini:2007cm,Sun:2014uq}; it is
not clear which will bring the greatest benefits, but either method would
further improve sensitivity. We can also incorporate automated energy
calibration \citep{Runkle:2009ev} and rain detection \citep{Mercier:2009gv} to
better estimate background spectra with less variance.


\section*{Acknowledgments}

This work would not be possible without the support of many people at the
University of Texas: Todd Hay and Patrick Vetter of Applied Research
Laboratories; Steven Biegalski of the Nuclear Engineering Teaching Laboratory;
and Scott Pennington of Environmental Health and Safety. Chad Schafer
contributed many useful suggestions and comments. Bridgeport Instruments
provided useful advice and technical support, and James Scott made many
statistical suggestions for early incarnations of the project.

\section*{References}

\bibliographystyle{elsarticle-num}
\bibliography{paper}

\end{document}